\documentclass{emulateapj}

\usepackage{natbib}
\usepackage{amssymb}
\usepackage[displaymath]{lineno}
\usepackage{graphicx}
\usepackage{subfigure}
\usepackage{lineno}

\usepackage{captcont}
\usepackage[final,              % override "draft" which means "do nothing"
        colorlinks,                     % rather than outlining them in boxes
        linktocpage,            % link just the pages, not the full titles
        linkcolor=blue,         % override truly awful colour choices
        citecolor=blue,         % (ditto)
        urlcolor=blue,          % (ditto)
        breaklinks=true,        % so that links wrap around to multiple lines if necessary
        ]{hyperref}
\usepackage[all]{hypcap} %needed to get figure links right

\shorttitle{Jet Dynamics and Calorimetry}
\shortauthors{Wygoda et al.}

\begin{document}

\title{Relativistic Jet Dynamics and Calorimetry of Gamma-Ray Bursts}

\author{
N.~Wygoda\altaffilmark{1,2},
E.~Waxman\altaffilmark{1}, and
D.~A.~Frail\altaffilmark{3}
}

\altaffiltext{1}{Department of Particle Physics \& Astrophysics,
  The Weizmann Institute of Science, Rehovot 76100, Israel}
\altaffiltext{2}{Department of Physics, NRCN, P.O. Box 9001, Beer-Sheva 84015, Israel}
\altaffiltext{3}{National Radio Astronomy Observatory, Array Operations Center, Socorro, NM 87801, USA}

\begin{abstract}

  We present numerical solutions of the 2D relativistic hydrodynamics
  equations describing the deceleration and expansion of highly
  relativistic conical jets, of opening angles $0.05\le\theta_0\le0.2$,
  propagating into a medium of uniform density. Jet evolution is followed from a
  collimated relativistic outflow through to the quasi-spherical
  non-relativistic phase. We show that relativistic sideways
  expansion becomes significant beyond the radius
  $r_\theta$ at which the expansion Lorentz factor drops to
  $\theta_0^{-1}$. This is consistent with simple analytic estimates,
  which predict faster sideways expansion than has been claimed based on
  earlier numerical modeling.
  For $t>t_s=r_\theta/c$ the emission of radiation from the jet blast
  wave is similar to that of a spherical blast wave carrying the same energy.
  Thus, the total (calorimetric) energy of GRB blast waves may be estimated
  with only a small fractional error based on $t>t_s$ observations.

\end{abstract}


\section{Introduction}
\label{sec:intro}

The dynamics of gamma-ray burst (GRB) jets and the spectral and temporal
evolution of their afterglows remain an important problem (see
\citealt{gr10} for a review). Much of what we know about GRB
progenitors and the central engines that power them comes from
multi-wavelength observations of their afterglows. From the analytic
models which predict the evolution of these light curves it is
possible to extract estimates of jet opening angles, the energetics of
the outflows, and the properties of the circumburst medium
\citep{pk02,cfh+10,yhsf03}. However, more recent numerical modeling
has claimed that there are strong discrepancies between the
analytic and numerical models of GRB jets.

A conical jet-like outflow expanding at a Lorentz factor $\Gamma$ evolves as if it
were a conical section of a spherical outflow as long as $\Gamma>\theta_0^{-1}$,
since for $\Gamma>\theta_0^{-1}$ the (rest frame) transverse light crossing time of the jet
is larger than the expansion/deceleration time. At this stage, the flow is described by
the spherical Blandford-McKee (BM) blast wave solutions \citep{bm76}, with $\Gamma^2r^3=17E_{\rm iso}/16\pi nm_pc^2$.
Here $r$ is the blast wave radius, $n$ is the ambient medium number density,
and $E_{\rm iso}$ is the isotropic equivalent energy,
related to the (two-sided) jet energy by $E_{\rm jet}=\frac{1}{2}\theta_0^2E_{\rm iso}$
(note that $\theta_0$ is the angular radius). For $\Gamma>\theta_0^{-1}$
a distant on axis observer cannot distinguish a jet from a sphere since the
emitted radiation is beamed into a $1/\Gamma$ cone.

The flow decelerates to $\Gamma=\theta_0^{-1}$ at source frame time
\begin{equation}
    t_{\theta}=r_\theta/c= 230 \left(E_{\rm iso,53}/n_0\right)^{1/3}\theta_{0,-1}^{2/3}\ {\rm day},
\label{eq:t_theta_burster} \end{equation}
corresponding to an observer's frame time
\begin{equation}
    t_{\theta,\oplus}=\frac{1}{4\Gamma^2}t_{\theta}
    =0.6\left({E_{\rm iso,53}\over n_0}\right)^{1/3}\theta_{0,-1}^{8/3}\ {\rm day},
\label{eq:t_theta_obs} \end{equation}
where $\theta_{0}=10^{-1}\theta_{0,-1}$, $E_{\rm iso}=10^{53}E_{\rm iso,53}$~erg
and $n=1n_0\rm cm^{-3}$. (For a burst located at redshift $z$, all observed
times should be increased by a factor $1+z$; we do not explicitly show this
correction in our eqs.) The sideways expansion is
expected to be relativistic as long as
the blast wave is relativistic and the post-shock energy density
is relativistic \citep{rho99}. If this is the case, at $t>t_\theta$
the lateral expansion rapidly increases the jet opening angle and
accelerates its deceleration \citep{rho99}, reducing $\Gamma$ to
$\sim1$ with only a logarithmic increase of $r$ (to $\sim\ln\theta_0^{-1}\times r_\theta$).
Thus, the observed time scale for the flow to become transrelativistic is \citep{lw00}
\begin{equation}
    t_{s,\oplus}\approx t_{\theta,\oplus}+ r_\theta/c\approx t_{\theta}
    = 230\left(E_{\rm iso,53}/n_0\right)^{1/3}\theta_{0,-1}^{2/3}\ {\rm day}.
\label{eq:t_s_burster}
\end{equation}
On a similar time scale, the flow is expected to become quasi-spherical, i.e.
the jet is expected to expand to $\theta\sim1$,and the outflow
is subsequently expected to evolve into the spherical non-relativistic
Sedov-von Neumann-Taylor (ST) flow.

This simple analytic description of jet expansion was challenged by a series of
numerical calculations \citep{gmp+01,cgv04,zm09,mk10}. It was argued, based on the
numerical results, that the sideways expansion of the jet is not relativistic,
and that the jet retains its narrow original opening angle, $\theta_0$,
as long as it is relativistic \citep{granot07,zm09}. This implies that the jet continues to evolve
like a conical section of a spherical outflow with energy $E_{\rm iso}$,
with Lorentz factor following the BM solution, up to the radius $r_{\rm NR}=ct_{\rm NR}$
at which it becomes sub-relativistic,
\begin{equation}
    t_{\rm NR}=\left({17E_{\rm iso}\over16\pi n m_pc^5}\right)^{1/3}
    =1100\left({E_{\rm iso,53}\over n_0}\right)^{1/3}\ {\rm day}.
\label{eq:t_NR_burster} \end{equation}

The different descriptions of jet evolution inferred from analytic and
numeric modeling lead to different predictions for the observed
properties of GRB afterglows \citep[e.g.][]{vlm+10}. For example,
the suppression of the observed flux produced by the jet blast wave
at $t>t_{\theta,\oplus}$, compared to that produced by a spherical
blast wave with the same $E_{\rm iso}$, is smaller (i.e. the "jet break"
is less pronounced) if the jet does not expand significantly while
it is relativistic. Furthermore, if the jet does not expand significantly
and remains highly collimated and relativistic at $t>t_s$, then the
accuracy of the late-time calorimetric estimates of the jet energy,
which assume quasi-spherical emission at $t\sim t_s$ \citep{fwk00,bkf04},
is questionable.

The main goal of this letter is to resolve the apparent discrepancy between the
analytic and numeric description of jet expansion. A more detailed discussion
of the properties of the flow and of the emitted radiation will be given in a
subsequent more comprehensive publication \citep{wygoda11b}. Our numerical
calculations are described
in \S~\ref{sec:model}, and their results regarding jet expansion are described in
\S~\ref{sec:dynamics}. In \S~\ref{sec:radiation} we briefly discuss the implications
for jet breaks and GRB calorimetry. Our conclusions are summarized in
\S~\ref{sec:discuss}.

\section {Numerical calculations}
\label{sec:model}

We use the RELDAFNA code \citep{klein10} to numerically solve the
2D special relativistic hydrodynamics equations describing the flow
of an ideal fluid with a constant polytropic index, $\gamma=4/3$.
RELDAFNA is a Godunov type Eulerian code, with second order accuracy
time and space integration. It uses adaptive mesh refinement
and is massively parallelized, allowing the use of effectively high
resolution even in multiscale problems such as the current jet
simulations. RELDAFNA was tested \citep[][see also \S~\ref{sec:dynamics}]{klein10}
by comparing its
solutions for standard test problems to those of similar codes
\citep{zm06, mkcg07}, and was shown to perform similarly.

The initial conditions chosen for our numerical calculations were a conical section of opening angle $\theta_0$ within which the flow fields are given by the BM solution for $E_{\rm iso}=10^{53}$~erg and initial density $n=1 \rm cm^{-3}$, surrounded by a static uniform cold gas of density $n=1 \rm cm^{-3}$ and pressure $p_0=10^{-10}nm_pc^2$. The radius of the conical section was chosen so that the Lorentz factor of the fluid behind the shock is $\Gamma=20$. We present solutions for $\theta_0=$0.2, 0.1 and 0.05 (corresponding to $E_{\rm jet}=2\cdot10^{51}, 5\cdot10^{50}$, and $1.25\cdot10^{50}$ erg). The $\theta_0=0.2$ simulation is similar to the simulation presented in \citep[][hereafter ZM09]{zm09}. The only difference is that we use $\gamma=4/3$, instead of an equation of state for which $\gamma$ varies smoothly between $\gamma=4/3$ for relativistic material and $\gamma=5/3$ for non-relativistic material. Our choice is inaccurate for late times, when the flow becomes non-relativistic, but this inaccuracy is not expected to affect our results qualitatively. Moreover, if a significant fraction of the post-shock energy density is carried by magnetic fields and relativistic electrons, as required in order to account for afterglow observations, $\gamma$ remains close to its relativistic value further into the non-relativistic flow stage.

The size of the finest numerical cells in the simulation was initially taken as $5.6\cdot10^{13}$~cm, similar to ZM09.
The results of our simulations were checked for convergence by increasing the grid resolution by a factor of 4 in each dimension. Increased resolution calculations were carried out both for initial conditions identical to those of the nominal calculations, and for initial conditions with a reduced radius of the conical section corresponding to a post-shock BM Lorentz factor of $\Gamma=40$. These convergence tests indicated that while the Lorentz factor behind the shock, as well as the lightcurves of the high frequencies that depend strongly on the high $\Gamma$ region, are not yet fully converged, the spreading of the jet is converged to a level of 10\%. For example, the time it takes the jet angle to double its initial value decreased in the convergence test by $\sim8\%$ for $\theta_0=0.2$ and by $\sim25\%$ for $\theta_0=0.05$. The results presented in the next sections are the ones obtained with the higher numerical resolution. A more detailed analysis of the numerics will be given in \citet{wygoda11b}.

\section {Jet expansion}
\label{sec:dynamics}

\begin{figure}[h]
\epsscale{1} \plotone{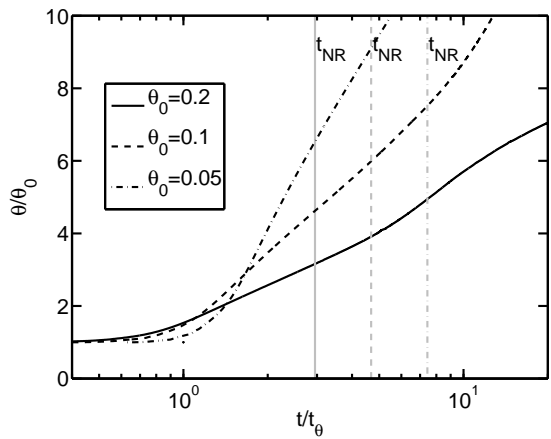}
\caption{The evolution of the jet opening angle, $\theta(t)/\theta_0$,
as a function of time (measured in the source frame and normalized to $t_\theta$), for $\theta_0=0.2$ (solid), 0.1 (dashed) and 0.05 (dash-dotted). $\theta$ is defined as the cone opening angle within which 95\% of
the energy, excluding rest mass energy, is included. The 3 lines denoted $t_{\rm NR}$ show $t_{\rm NR}/t_\theta$ for the 3 values of $\theta_0$.
\label{fig:2djet_theta02_rshock}}
\end{figure}

Figure~\ref{fig:2djet_theta02_rshock} presents the evolution of
the jet opening angle, $\theta(t)$, as a function of time.
$\theta$ is defined as the cone opening angle within which 95\% of
the energy, excluding rest mass energy, is included. We find that
significant sideways expansion begins at $t\sim t_\theta\ll t_{\rm NR}$,
in accordance with the analytic estimates described in \S~\ref{sec:intro}.
Narrower jets begin expanding earlier, $t_\theta\propto\theta^{2/3}$,
in accordance with the analytic estimates, and are therefore expected
to also decelerate earlier.

\begin{figure}[h]
\epsscale{1} \plotone{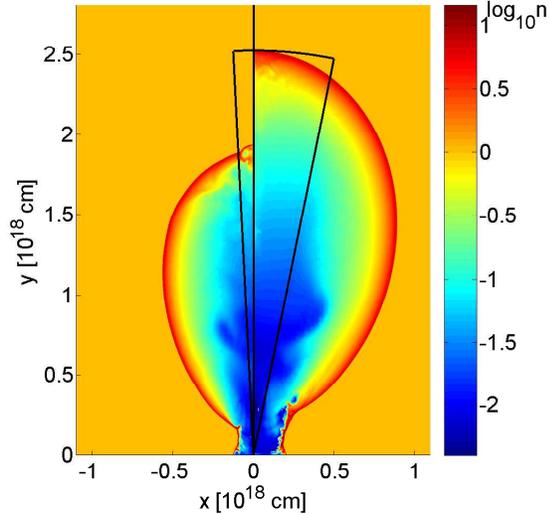}
\caption{The density distribution of the $\theta_0=$~0.2 (right) and 0.05 (left) jet flows at identical time $t=0.95 t_{\rm NR}$ ($t_{\rm NR}$ is the same for both jets since $E_{\rm iso}$ and $n$ are identical for both). The black lines
indicate the shape and size the jet would have reached if it behaved like a conical section of a sphere.
\label{fig:large_vs_small_jet}}
\end{figure}

The latter point is demonstrated
in figure~\ref{fig:large_vs_small_jet}, which shows the density distribution of the
$\theta_0=$~0.2 and 0.05 jet flows at identical time $t=0.95t_{\rm NR}$
(note that $t_{\rm NR}$ is independent of $\theta_0$). The $\theta_0=0.2$ jet has tripled
its opening angle and its tip is still close to its ``isotropic equivalent
location'', i.e. the location of a spherical blast wave with the same $E_{\rm iso}$.
The opening angle of the $\theta_0=0.05$ jet has increased by more than an
order of magnitude, and its spreading has significantly slowed down its expansion.
The influence of jet expansion at $t_{\rm NR}$ is much stronger for the narrower
jet, in accordance with the analytic analysis described in \S~\ref{sec:intro}:
The ratio of $t_{\rm NR}$ to $t_s$ is close to unity for the $\theta=0.2$ jet,
and significantly larger for the $\theta=0.05$ jet: at the source
frame $t_s\approx t_\theta+r_\theta/c=2t_\theta$ so that
$t_{\rm NR}/t_s\approx1/2\theta_0^{2/3}=1.5\ ,3.7$ for $\theta_0=0.2\ ,0.05$.

We thus find that the jet sideways expansion is relativistic and becomes
significant at $t\sim t_\theta\propto\theta_0^{2/3}$, and that this expansion
leads to deceleration to sub-relativistic velocity on a time scale
$t_s\propto\theta_0^{2/3}$, which for $\theta_0\ll1$ is much smaller than
$t_{\rm NR}\propto\theta_0^0$. This behavior is consistent
with the analytic analysis, and inconsistent with the claims based on
earlier numerical modeling, that jet expansion is not significant up
to $t\sim t_{\rm NR}$.

\begin{figure}[h]
\epsscale{1} \plotone{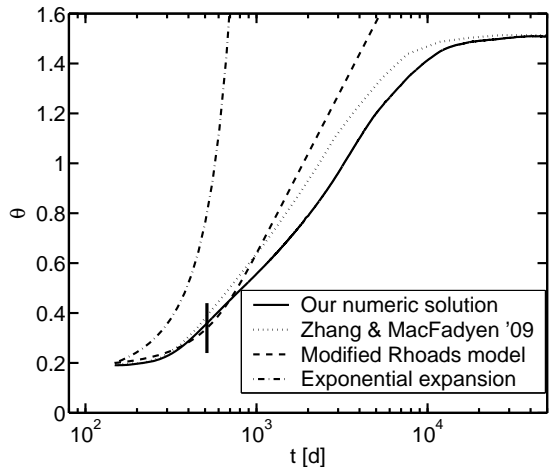}
\caption{Comparison of $\theta$ as function of source frame time obtained for the $\theta_0=0.2$ jet in our numerical calculation (solid line), with that obtained by ZM09 (dotted), and by the modified Rhoads' model (dashed, eq.~\ref{eq:rhoads}). The short vertical line denotes the time at which $\Gamma=2$, up to which Rhoads' model applies. The dash-dotted line shows the exponential, $\ln\theta/\theta_0\propto t$, to which the numerical results were compared in ZM09 (see their fig.~3).
%\textbf{Change xlabel to $t$~[d], ylabel to $\theta$, "Simulation" to "Our numeric solution", "Rhoads exponential..." to "Exponential expansion".}
\label{fig:2djet_theta02}}
\end{figure}

In order to identify the origin of this apparent discrepancy, we compare our numerical
results to those of ZM09 in fig.~\ref{fig:2djet_theta02}.
The figure demonstrates that the jet expansion obtained in our calculation is
similar to that obtained in ZM09. The conclusion that sideways expansion is not
relativistic, and unimportant until $t\sim t_{\rm NR}$, was reached in ZM09 based
on noting that the growth of $\theta(t)$ is much slower than the exponential growth
expected for relativistic sideways expansion (see fig.~\ref{fig:2djet_theta02}).
This conclusion is, however, not valid, since exponential growth
is expected only for $\theta_0^{-1}\gg\Gamma\gg1$, and is not
applicable for the evolution of the $\theta_0=0.2$ jet under consideration,
for which expansion becomes significant only for $\Gamma<\theta_0^{-1}=5$.

For this
regime of $\Gamma$, one cannot use the exponential approximation, but should rather
solve the differential equation
\begin{equation}\label{eq:rhoads}
    d\theta/dr=c_s/{\Gamma c r},
\end{equation}
describing relativistic sideways expansion at the post shock speed of sound $c_s$
in the jet frame ($c_s=c/\sqrt{3}$ for $\Gamma\gg1$),
along with mass and momentum conservation, that determine $\Gamma(r,\theta)$
\citep[for more details see][]{rho99}. Note that we replace eq.~(3) of \citet{rho99}
with the more accurate eq.~(\ref{eq:rhoads})
\citep[see also][]{piran00}. This modification leads to a significant
modification of $\theta(r)$ only for $\theta>0.4$. For $\theta_0^{-1}\gg\Gamma\gg1$, the solution
of eq.~(\ref{eq:rhoads}) is indeed exponential, $\ln\theta/\theta_0\propto r^{3/2}$.
However, such a regime does not exist for $\theta_0=0.2$. As demonstrated in the figure,
the numerical results are in good agreement with the solution of the simple model
of eq.~(\ref{eq:rhoads}), and therefore
confirm the validity of the analytic estimates described in \S~\ref{sec:intro}.

\section {Light Curves And Calorimetry}
\label{sec:radiation}

We have calculated the synchrotron emission expected to be produced by shock accelerated electrons assuming that the magnetic field and the electrons hold a constant fraction $\epsilon_e=\epsilon_B=0.1$ of the internal energy, and that the electron energy distribution is a power law with index $p=2.4$. Electron cooling and synchrotron self absorption are neglected.% \textbf{Give here a brief description of the assumptions, including no self-absorption.}

\begin{figure}[h]
\epsscale{1} \plotone{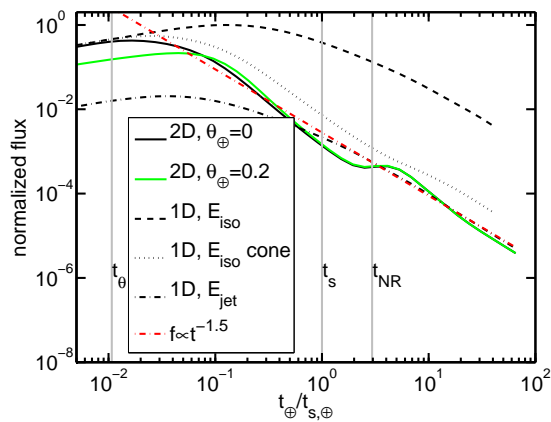}
\caption{The (2D) $\theta_0=0.2$ jet radio lightcurve ($\nu\approx3$~GHz) for observers lying on-axis (solid black line) and at $\theta_\oplus=0.2$ rad, compared with those of several spherical (1D) fireballs: $E=E_{\rm iso}$ (dashed), $E=E_{\rm iso}$ with emission only from conical section (dotted), and $E=E_{\rm jet}$ (dash-dotted). The red line denotes the asymptotic ST behavior. \label{fig:2D_lightcurves}}
\end{figure}

Figure~\ref{fig:2D_lightcurves} shows radio light-curves ($\nu\approx$~3GHz, for which self-absorption is not important) predicted by the numerical model for the $\theta_0=0.2$ jet, for observers lying on the jet axis and at an angle $\theta_{\rm obs.}=\theta_0$. The numerical lightcurves of the jet are compared with those predicted for spherical (1D) fireballs with total energy $E_{\rm iso}$ and $E_{\rm jet}$, as well as with that predicted for a conical section with opening angle $\theta_0$ of a spherical $E_{\rm iso}$ fireball, representing the lightcurve predicted for a non-expanding jet. Also shown is the radio flux of a spherical fireball with energy  $E_{\rm jet}$, assuming its evolution is described by the non-relativistic Sedov-von Neumann-Taylor solution.

The jet emission is suppressed, compared to that of a spherical blast wave with energy $E_{\rm iso}$, at $t>t_{\theta}$. The suppression is larger than would be predicted for a non-expanding jet \citep[i.e. not due to the ``missing flux'' from the absent $\theta>\theta_0$ parts of the shell, but rather to the jet spreading, in accordance with][]{vmwk10}.
The figure also demonstrates that at $t>t_s$, the observed flux is similar to that of a spherical fireball with energy $E_{\rm jet}$, for observers lying both on-axis and at an angle $\theta=\theta_0$. Moreover, although at $t\sim t_s$ the jet has not yet reached full spherical symmetry and is still mildly collimated (see fig.~\ref{fig:2djet_theta02_rshock}), at $t>t_s$ the flux is well approximated by that of a blastwave following the non-relativistic ST evolution.

In Figure \ref {fig:2D_lightcurves_frel} we compare the $\theta_0=0.2, 0.1, 0.05$ jet lightcurves to those of spherical (1D) fireballs with the corresponding $E=E_{\rm jet}$. Scaling $t_\oplus$ with $t_{s,\oplus}$ brings the lightcurves to a similar (universal) form, implying that significant jet spreading occurs at $t\sim t_\theta\propto\theta^{2/3}$. Although, as noted above, at $t\sim t_s$ the jet does not yet reach full spherical symmetry, from $\sim 0.3t_s$ onward the lightcurves do not depart from those of the 1D $E=E_{\rm jet}$ fireballs by more than $\sim50\%$. A more detailed analysis of the jet lightcurves will be presented in \citet{wygoda11b}.

\section{Discussion and Conclusions}
\label{sec:discuss}

While there still remain many unsolved problems regarding the
structure and dynamics of GRB jets that can only be addressed by
detailed hydrodynamic modeling, our work has demonstrated that
analytic estimates provide a reasonable description of the
behavior of the jet and the evolution of its afterglow.

We have shown that relativistic sideways expansion becomes significant
at $t>t_\theta\propto\theta_0^{2/3}$ (fig.~\ref{fig:2djet_theta02_rshock}),
in accordance with analytic
estimates, and that the expansion is well described by the modified Rhoads
model (fig.~\ref{fig:2djet_theta02}, eq.~\ref{eq:rhoads})).
Our numerical results are consistent with those of ZM09, who calculated
$\theta_0=0.2$ jet evolution. The apparent discrepancy between earlier
numerical and analytic results arose because the simulations weren't
compared to the full solution of Rhoads' model
(fig.~\ref{fig:2djet_theta02}, \S~\ref{sec:dynamics}), and because
for the large $\theta_0$ chosen it is difficult to test the relativistic
expansion assumption since $t_{\rm NR}$ and $t_s$ are similar,
$t_{\rm NR}/t_s\approx1/2\theta_0^{2/3}=1.5$
(see \S~\ref{sec:dynamics}).

\begin{figure}[h]
\epsscale{1} \plotone{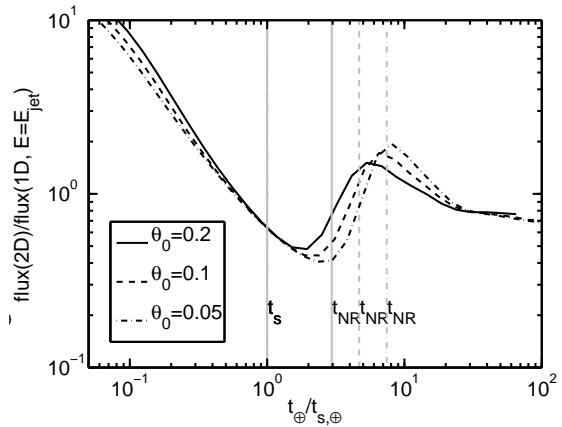}
\caption{Ratio between the 2D jet flux and that of a 1D spherical fireball with $E=E_{\rm jet}$ for $\theta_0=0.2$ (solid), 0.1 (dashed) and 0.05 (dash-dotted).\label{fig:2D_lightcurves_frel}}
\end{figure}

Jet expansion has a significant effect on its observed properties.
The suppression of the flux at $t>t_\theta$ is stronger than
in the absence of spreading (fig.~\ref{fig:2D_lightcurves}), and at
$t>t_s$ the emission of radiation from the jet blast wave is similar
to that of a spherical blast wave carrying the same energy
(fig.~\ref{fig:2D_lightcurves},\ref{fig:2D_lightcurves_frel}). Moreover,
although at $t\sim t_s$ the jet has not yet reached full
spherical symmetry and is still mildly collimated
(see fig.~\ref{fig:2djet_theta02_rshock}),
at $t>t_s$ the flux is well approximated by that of a blastwave
following the non-relativistic ST evolution (fig.~\ref{fig:2D_lightcurves}).
Thus, the total (calorimetric) energy of GRB blast waves may be estimated
with only a small fractional error based on $t>t_s$ observations. We expect
to see this technique and its variants \citep{fwk00,vkr+08, sb11} applied to
increasing numbers of GRB afterglows when the new generation of
facilities (EVLA, LOFAR) starts full operation.

\acknowledgments
We thank Y. Elbaz and Y. Klein for their authorization to use the RELDAFNA code, as well as for useful discussions.
This research was partially supported by Minerva, ISF and Israel's Universities Planning and Budgeting committee grants.
% NRAO will pay the most of the page charges if this tag line is in the paper.
The National Radio Astronomy Observatory is a facility of the
National Science Foundation operated under cooperative agreement by
Associated Universities, Inc.

\end{document}